\documentclass[preprint2,12pt]{emulateapj}
\usepackage{amsmath}
\usepackage{natbib}

\newcommand{\G}{\Gamma}

\providecommand {\rce  }{$R_\mathrm{ce}\,$}
\providecommand {\bapp }{$\beta_\mathrm{app}\,$}
\providecommand {\bappmax}{$\beta_\mathrm{app,max}\,$}
\providecommand {\bappns}{$\beta_\mathrm{app}$}

\providecommand {\lgls }{$R_\mathrm{p}\,$}
\providecommand {\lp   }{$L_\mathrm{p,syn}\,$}
\providecommand {\vp   }{$\nu_\mathrm{p,syn}\,$}
\providecommand {\lkin }{$L_\mathrm{kin}\,$}
\providecommand {\rcens  }{$R_\mathrm{ce}$}

\providecommand {\lglsns }{$R_\mathrm{p}$}
\providecommand {\lpns   }{$L_\mathrm{p,syn}$}
\providecommand {\vpns   }{$\nu_\mathrm{p,syn}$}
\providecommand {\lkinns }{$L_\mathrm{kin}$}
\providecommand {\lpicns   }{$L_{\mathrm{p,IC}}$}
\providecommand {\vpicns   }{$\nu_{\mathrm{p,IC}}$}
\providecommand {\lpic}{$L_{\mathrm{p,IC}}\,$}
\providecommand {\vpic}{$\nu_{\mathrm{p,IC}}\,$}

\slugcomment{Accepted for Publication in the Astrophysical Journal Letters}

\shorttitle{External Compton Emission in High-Power Blazars}
\shortauthors{Meyer, Fossati, Georganopoulos, Lister}

\begin{document}

\title{Collective Evidence for Inverse Compton emission from External Photons in High-Power Blazars}

\author{Eileen T. Meyer\altaffilmark{1}, Giovanni Fossati \altaffilmark{1}, Markos Georganopoulos\altaffilmark{2,3},  Matthew L. Lister \altaffilmark{4}}
 
\altaffiltext{1}{Department of Physics and Astronomy, Rice University, Houston, TX 77005}
\altaffiltext{2}{Department of Physics, Joint Center for Astrophysics,
University of Maryland Baltimore County, 1000 Hilltop Circle,
Baltimore, MD 21250, USA}
\altaffiltext{3}{NASA Goddard Space Flight Center, Code 660, Greenbelt, MD 20771, USA}
\altaffiltext{4} {Department of Physics, Purdue University, 525 Northwestern Ave., West Lafayette, IN 47907, USA}

\begin{abstract}
 We present the first collective evidence that \emph{Fermi}-detected
 jets of high kinetic power (\lkinns) are dominated by inverse Compton
 emission from upscattered external photons. Using a sample with a
 broad range in orientation angle, including radio galaxies and
 blazars, we find that very high power sources ($L_\mathrm{kin} >$
 10$^{45.5}$ erg s$^{-1}$) show a significant increase in the ratio of
 inverse Compton to synchrotron power (Compton dominance) with
 decreasing orientation angle, as measured by the radio core dominance
 and confirmed by the distribution of superluminal speeds. This
 increase is consistent with beaming expectations for external Compton
 (EC) emission, but not for synchrotron-self Compton (SSC) emission.
 For the lowest power jets ($L_\mathrm{kin} <$ 10$^{43.5}$ erg
 s$^{-1}$), no trend between Compton and radio core dominance is
 found, consistent with SSC.  Importantly, the EC trend is not seen
 for moderately high power
flat spectrum radio quasars
with strong external photon fields. Coupled with the evidence that jet
power is linked to the jet speed \citep{kha10}, this finding suggests
that external photon fields become the dominant source of seed photons
in the jet comoving frame only for the faster and therefore more
powerful jets.
\end{abstract}


\section{Introduction \label{section:intro}}
The relativistic jets of radio-loud active galactic nuclei (AGN) are
copious gamma-ray emitters, as was first discovered by \emph{EGRET}
\citep{har92}, and confirmed by the \emph{Fermi} large area telescope
(LAT), which has associated over 800 sources with radio-loud AGN in
the second catalog \citep[2LAC;][]{ack11_2lac}. Most of these are
blazars, seen with the jet axis along the line of sight, though
\emph{Fermi} has also detected the jets of several radio galaxies
\citep[RG;][]{abd09_fermi_misaligned_agn,kat11_blrg}, which are
misaligned blazars under the standard unification scheme
\citep{urr95}.

The lower-energy peak in the jet spectrum is well-understood as
synchrotron emission from relativistic electrons in the jet.  The
high-energy component, peaking from X-ray to TeV energies, is
attributed to photons upscattered by the same relativistic electrons
to higher energies via the inverse Compton (IC) process (see
\citealt{bot07} for a review). These photons could arise from the jet
synchrotron emission \citep[synchrotron self-Compton emission,
  SSC;][]{mar92,mar96} or from external sources such as the accretion
disk \citep[][]{der92}, broad-line region
\citep[BLR;][]{sik94}, or molecular torus \citep[MT;][]{bla00,sik09},
i.e. external Compton (EC) emission. Identifying the IC emission
mechanism is a diagnostic for the location of the gamma-ray
emitting region, a currently open issue \citep[e.g.][]{agu11}.  However, the
spectral energy distributions (SEDs) of individual sources are rarely
sufficiently constraining of the IC mechanism due to the number of
free parameters \citep[][]{sik97}.

The beaming pattern (how apparent luminosity changes with
orientation) is different for EC and SSC emission. Thus,
a \emph{collective} study using a sample of sources at different
orientations can be used to identify the gamma-ray emission mechanism.
In this letter we discuss the effect of jet power and orientation on
the observed IC power in view of the recent suggestion of a dichotomy
in radio-loud AGN (\S \ref{icenv}). We then show that the collective
beaming pattern for a sub-set of high-power sources appears to support
EC models rather than SSC for this population (\S \ref{echp}), and
discuss the implications for the gamma-ray emission region.

\section{A Dichotomy in Radio-loud AGN}
\label{icenv}

\subsection{The Synchrotron Plane}
\label{benv}

Using a large sample of jets, we recently found evidence that
radio-loud AGN form two populations in the plane of synchrotron peak
luminosity (\lpns) versus peak frequency \citep[\vpns;][hereafter
  M11]{mey11}. A population of `weak' jets consists of sources with
low \lp which appear to be most aligned at high \vp ($\sim$10$^{17}$
Hz) and trace out a shallow track on the plane as they become less
aligned (dropping more in \vp than \lpns). A separate population of
`strong' jets with higher \lp and \vp $\lesssim$ 10$^{15}$ Hz appear
to drop rapidly in luminosity with misalignment (as measured by radio
core dominance). Importantly, the weak jets exhibit jet kinetic powers
below 10$^{44.5}$ erg s$^{-1}$ while all sources with \lkin $>$
10$^{44.5}$ erg s$^{-1}$ are on the strong-jet branch. The weak/strong
divide in the synchrotron plane, which we associate with the
morphological dichotomy in Fanaroff-Riley (FR) type 1 and 2 RG, may
also be mapped to a critical transition in accretion efficiency
\citep{ghi01,ghi09_fermi_divide,geo11}.

\begin{figure}
  \includegraphics[width=1.0\linewidth]{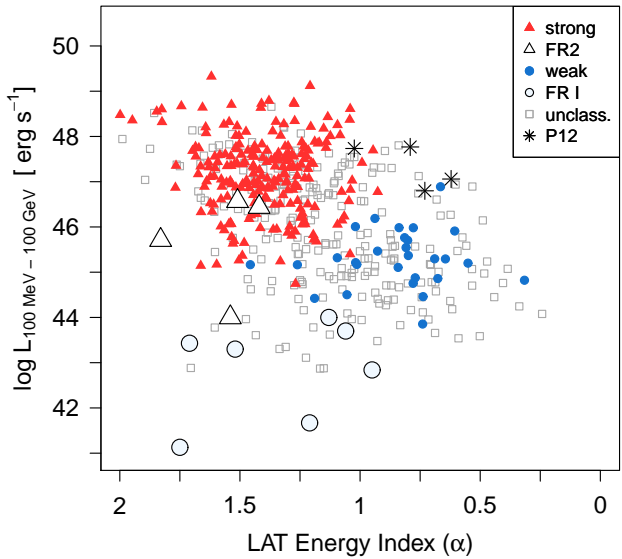}
  \caption{ This plot of total LAT-band luminosity versus gamma-ray
    energy index $\alpha$ (approximating \lpic versus \vpicns) is the
    high-energy analog of the synchrotron plane discussed in
    M11. Blazars have been divided based on the luminosity and
    location of their synchrotron peaks into strong-jet and weak-jet
    blazars as described in the text. Those without \vp and \lp are
    `unclassified'. FR2 RG appear to have similar \vp (similar
    $\alpha$) to the strong-jet blazars while FR1 RG have much lower
    values of \vp (higher $\alpha$) relative to the weak-jet blazars
    with which they are associated. The FR1 would be detectable by
    \emph{Fermi} even at much harder $\alpha$
    \citep[see][Figure~15]{ack11_2lac}. The region at upper right
    remains empty despite recent claims of high-power, high-peak
    sources \citep[shown as black stars,][]{pad12}, matching what has
    been found in the synchrotron plane.}
    \label{fermienv}
\end{figure}

\subsection{The Blazar Envelope at High Energies}
In light of the dichotomy discussed above, we present in
Figure~\ref{fermienv} the total LAT-band
luminosity\footnote{$L_\mathrm{100 MeV - 100 GeV}$ = 4$\pi d_L^2
  S_\gamma/(1+z)^{1-\alpha}$ for power-law sources, where $S_\gamma$
  is the catalog integrated energy flux and $\alpha$ = $\Gamma-$1
  using the published photon index $\Gamma$. For sources fit with
  Log-Parabolic spectra, the integrated luminosity is calculated from
  the 2LAC fitted values with a k-correction.} (a proxy for the IC
peak luminosity, \lpicns), versus the LAT-band energy index $\alpha$
(a proxy for peak frequency, \vpicns) for the entire
\emph{Fermi}-detected sample of radio-loud AGN of known redshift (data
taken from the 2LAC). Sources with higher \vpic will have harder
(smaller) values of $\alpha$, thus in the figure \vpic increases to
the right.

An empty region at upper right appears analogous to that seen in the
synchrotron plane. \cite{gio12} have suggested that this is a
selection effect due to a lack of redshifts for high-frequency-peaked,
high-luminosity sources, and \cite{pad12} discuss four such
candidates. However, these candidates do not cross into the
upper-right region in Figure~\ref{fermienv} (black stars). These
interesting sources appear to exhibit \vp at or above $\sim$ 10$^{15}$
Hz; however, the SED sampling is sparse, and it is difficult to rule
out an alternative explanation for the soft X-ray spectra such as an
extra emission component \citep[see e.g., the case of BL
  Lacertae;][]{rai10}. Importantly, for the several hundred
\emph{Fermi} sources lacking identifications or redshifts, nearly all
($>$99\%) have soft spectra ($\alpha>$ 1) or would require redshifts
$>$ 2 or higher to cross into this region, making it unlikely to be
empty due to selection effects.

The blazars in Figure~\ref{fermienv} have been divided in an
approximate way based on their synchrotron spectra into strong and
weak jets. We take the former to be those defined by \lp $>$ 10$^{45}$
erg s$^{-1}$, \vp $<$ 10$^{15}$ Hz, as well as flat spectrum radio quasar
(FSRQ) type sources outside this area. For the weak jets, we take all
sources outside this area, minus any FSRQ. While very rare cases of
FR1-like (e.g., weak) FSRQs exist, their occurrence is negligible in
the bright \emph{Fermi-}detected population considered here \citep[see
  e.g.,][and references therein]{kha10}. The eleven
\emph{Fermi}-detected RG are also shown \citep[data
  from][]{abd09_fermi_misaligned_agn}. The FR1 in
Figure~\ref{fermienv} have much lower \vpic (larger $\alpha$) compared
with the weak-jet sources, while there is little difference between
$\alpha$ for FR2 and strong jets, tentatively matching the different
misalignment paths for these populations that were found in the
synchrotron plane.

\subsection{The Importance of Jet Kinetic Power}
\label{sec:lkin}
In M11, we showed that \lkin is an important parameter in classifying
RL AGN. We have selected from the 2LAC a subset of 152 blazars (as
well as the detected radio galaxies) with estimates of the 300 MHz
isotropic lobe emission, which is scaled to estimate \lkin as in
M11. For 46 sources with adequate coverage of the high-energy SED, we
estimated \lpic from a two-sided parabolic fit to the 2LAC data in
combination with X-ray data taken from NED\footnote{NASA Extragalactic
  Database: http://ned.ipac.caltech.edu/} and/or \emph{Swift}/BAT
\citep{cus10_bat3}. For an additional 106 sources, \lpic is scaled
from the rest-frame luminosity at 1 GeV estimated from the 2LAC fitted
spectrum: log \lpic = log $L_\mathrm{1 GeV}$ + $\eta(\alpha-1)^2$, where
$\eta$=1.5 if $\alpha > 1$, $\eta=5$ otherwise. The difference
between the two estimates is typically less than 0.2 decades in log
$L$.

We use the radio core dominance \rcens $\equiv$ log
($L_\mathrm{core}$/$L_\mathrm{ext}$) as a tracer of the orientation
angle, where the core luminosity is measured at 1.4 GHz and the
extended at 300 MHz (see M11). As discussed in M11, the absolute
normalization between orientation angle and \rce depends on \lkinns,
with \rce decreasing with increasing orientation angle for a given
\lkinns.

In Figure~\ref{lgvsr} we show the 145 sources which have a known
redshift. \lpic is positively correlated with both \rce and \lkinns.
The OLS bisector fits to the combined blazar-RG sample shown have
slopes of 1.1, 1.5, 1.4, 1.3, 1.5 ($\pm$0.1) from lowest to highest
bin in \lkinns. The correlation between \lpic and \rce in each group
is significant and positive (R values from 0.37 to 0.89); the apparent
gamma-ray output of a blazar is therefore a strong function of both
\lkin and the orientation angle.

\subsection{Emission Mechanisms and Their Beaming Patterns}
\label{sec:mech}
For both synchrotron and SSC, the beaming pattern is
$L=L^\prime\delta^{p+\alpha}$ \citep{der95}, where $L$ assumes
isotropic emission in the galaxy frame, $L^\prime$ is the solid-angle
integrated luminosity in the comoving jet frame, $\delta$ is the
Doppler factor, and $\alpha$ the spectral index. For
the exponent p+$\alpha$, p=3 for a `moving blob', or p=2 for a
stationary feature in a continuous jet \citep{lin85}, with the
possibility of different values for different emitting regions. For
the EC case, the pattern is $L_\mathrm{EC} =
L_\mathrm{EC}^\prime\delta^{p+1+2\alpha}$ \citep{der95,geo01}.

\begin{deluxetable}{ccccccc}
\tabletypesize{\small}
\tablecolumns{7}
\tablewidth{0pt}
\tablecaption{Predicted Correlation Slope Values}
\tablehead{
  \multicolumn{3}{c}{index values} & \multicolumn{2}{c}{SSC} & \multicolumn{2}{c}{EC}\\
  $p_r$ & $p_\mathrm{syn}$ & $p_\gamma$ & $L_\gamma-R_\mathrm{ce}$ & $R_\mathrm{p}-R_\mathrm{ce}$ & $L_\gamma-R_\mathrm{ce}$ & $R_\mathrm{p}-R_\mathrm{ce}$\\
}
\startdata
3 & 3 & 3 & 1.1$-$1.3 & 0 & 1.7$-$2\phantom{.0} & 0.6$-$0.7\\
2 & 3 & 3 & 1.6$-$2\phantom{.0} & 0 & 2.4$-$3\phantom{.0} & 0.8$-$1\phantom{.0}\\
2 & 2 & 3 & 1.6$-$2\phantom{.0} & 0 & 2.4$-$3\phantom{.0} & 1.2$-$1.5\\
2 & 2 & 2 & 1.2$-$1.5 & 0 & 1.7$-$2.5 & 0.8$-$1\phantom{.0}\\
\enddata
\label{table}
\end{deluxetable}
\vspace{10pt}

The slope of \lpic versus \rce depends on the gamma-ray emission
process.  From the synchrotron beaming,
\begin{equation}
R_{ce} \equiv \log(L_\mathrm{core}/L_\mathrm{ext}) = \left(p_r+\alpha_r\right)\,\mathrm{log}\delta + c_1,
\label{eqrce}
\end{equation}
where the factors $c_n$ depend on the unbeamed luminosity and jet
power, but do not affect the slope.  We take $\alpha$ = 1 for the
peaks, and eliminating $\delta$ we obtain:
\begin{equation}
\log L_\mathrm{peak} = \left(\frac{b}{p_r+\alpha_r}\right)\,R_{ce}+c_2,
\end{equation}
where b = 1 + $p_\gamma$ (SSC) or b = 3 + $p_\gamma$ (EC). For
reference we list in Table~\ref{table} the expected slopes for various
cases of $p$ values assuming typical values of $\alpha_r$
(0$-$0.5). The slopes in Figure~\ref{lgvsr} appear to be more
consistent with SSC under a $p_\gamma$, $p_r$=3 or $p_\gamma$, $p_r$=2
scenario. However, the lack of \emph{Fermi}-detected misaligned FR2
sources (particularly, with \rce $< -$0.5) is likely to affect the
slope for the two highest bins in \lkinns in Figure~\ref{lgvsr}; we
show for reference the upper limits to \lpic\footnote{From the flux
  limit ($>$ 100 MeV): log $f_{100}$ = $-$8.3 + (5/6)($\alpha -$1.2),
  with average $\alpha=$1.3 for our sample and \lpic scaled from
  $L_\mathrm{1 GeV}$ (Section~\ref{sec:lkin}).}  for four RG in the
M11 sample. We take the rightmost of these points as the most
constraining: including it in the fit for the log \lkin = 45$-$45.5
bin increases the slope to 1.5$\pm$0.2, and for the highest bin to
1.7$\pm$0.3. These slopes, and our assumption that the Lorentz factor
($\Gamma$) of the gamma-ray emitting plasma is similar to that in
radio, are discussed below.

\section{The Compton Dominance in \emph{Fermi} Blazars}
\label{echp}

\subsection{EC in powerful FSRQ}
For sources with a high-energy component dominated by SSC, the Compton
dominance (\lgls) should remain constant over all orientations (no
correlation between \lgls and \rce expected). For the EC case,
\begin{equation}
R_p \equiv log(L_\mathrm{EC}/L_S) = \left(p_\gamma - p_\mathrm{syn}+2\right)\,\mathrm{log}\delta + c_3.
\label{eqrp}
\end{equation}  
Using Equation~\ref{eqrce} we have the relation
\begin{equation}
  R_p = \left(\frac{p_\gamma - p_\mathrm{syn}+2}{p_r+\alpha_r}\right)R_{ce} + c_4.
  \label{eqrprce}
\end{equation}
As shown in Table~\ref{table}, we expect a correlation with a slope
from 0.6$-$1.5 under the simplest assumptions for sources emitting
gamma-rays by EC.

\begin{figure}
\includegraphics[width=1.0\linewidth]{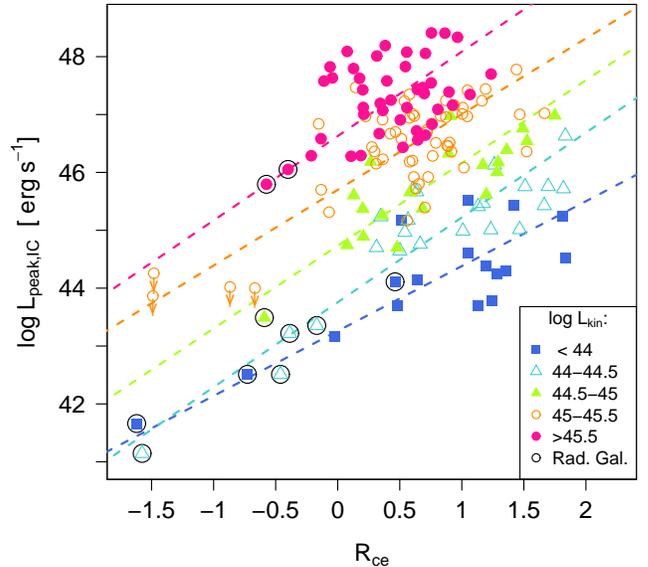}
\caption{Estimated IC peak luminosity (\lpicns) versus radio core
  dominance (\rcens) for 145 blazars and ten RG (circled) detected by
  \emph{Fermi}. The correlation between \lpic and \rce becomes clear
  when binning by \lkin. OLS bisector fits to sources grouped by \lkin
  have slopes 1.1, 1.5, 1.4, 1.3, 1.5 ($\pm$1; Pearson's R values
  0.81, 0.89, 0.82, 0.37, 0.44) beginning with the lowest bin. For an
  EC emission model, these slopes are predicted to be higher ($>
  1.7$); however, for the high-power sources, the lack of \emph{Fermi}
  detections for misaligned (i.e., low \rcens) sources will affect the
  apparent slope.}
\label{lgvsr}
\end{figure}

\begin{figure}
\includegraphics[width=1.0\linewidth]{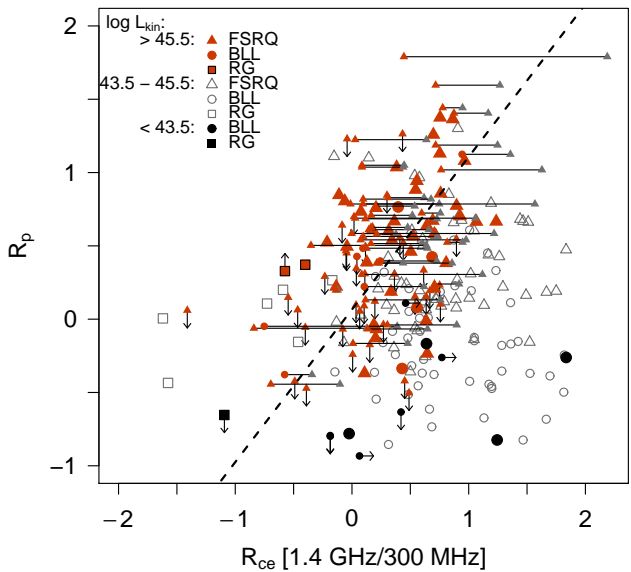}
\caption{The Compton dominance (\lglsns) versus radio core dominance
  (\rcens). A positive correlation is seen for the most
  powerful sources (shaded), as predicted for EC emission. The slope
  of the OLS bisector fit to the shaded points is 1.1$\pm$0.1, though
  the presence of the upper limits on \lgls and lower limits on \rce
  (range shown by black lines) makes this a lower limit on the real
  slope. Weak jets (black points) as well as moderately powerful FSRQ
  (open triangles) show no trend, as expected for simple SSC.
\vspace{20pt}
}
\label{ecsj}
\end{figure}

Figure \ref{ecsj} shows \lgls versus \rce for the sample discussed in
Section~\ref{sec:lkin}. Sources are divided into broad bins of
\lkinns. These bins include a mix of sources in terms of optical type
(FSRQ or BL Lac); however \emph{all} the very high power (VHP, \lkin
$>$ 10$^{45.5}$) are `strong-jet' types in our classification due to
their \lkin and position in the synchrotron plane, including many
\emph{apparent} BL Lacs, which likely suffer dilution of their broad
lines by the jet emission \citep[][]{geo98_thesis,ghi11_divide}. For
the moderate-power sources, the division into strong/weak roughly
follows the FSRQ/BL Lac divide.

No trend between \lgls and \rce is evident for blazars overall.
However, when VHP sources are selected, a positive correlation
emerges. We also show the upper limits
on \lgls for VHP sources from the sample of M11 not detected by
\emph{Fermi}\footnote{See previous footnote, with index $\alpha$
  estimated from the relation for detected sources:
  $\alpha$=$-$0.2$\times$\vp + 3.9}. VHP sources from the 2LAC and M11
with upper limits on $L_\mathrm{ext}$\footnote{Estimated at 300 MHz
  from the lowest-frequency SED point with spectral index
  $\alpha=1.2$.} are also shown; because of the binning on \lkin which
is scaled from $L_\mathrm{ext}$, we also show as connected gray points
the maximum \rce such that \lkin $>$ 10$^{45.5}$ erg s$^{-1}$.  The
OLS bisector fit through all VHP points (including upper/lower limits)
gives a slope of 1.1$\pm$0.1; this is a lower limit since most of the
\lgls upper limits are on the lower half of the correlation. A
sub-sample of sources with \rce values contemporaneous to
\emph{Fermi}, calculated from the average 15 GHz\footnote{scaled to
  1.4 GHz by $-$1.2 in log L} flux over the time-frame of the 2LAC as
measured by the Owens Valley Radio Observatory \citep[OVRO,][]{ric11},
gives an identical slope.

\vspace{10pt}
\begin{figure}
\includegraphics[width=1.0\linewidth]{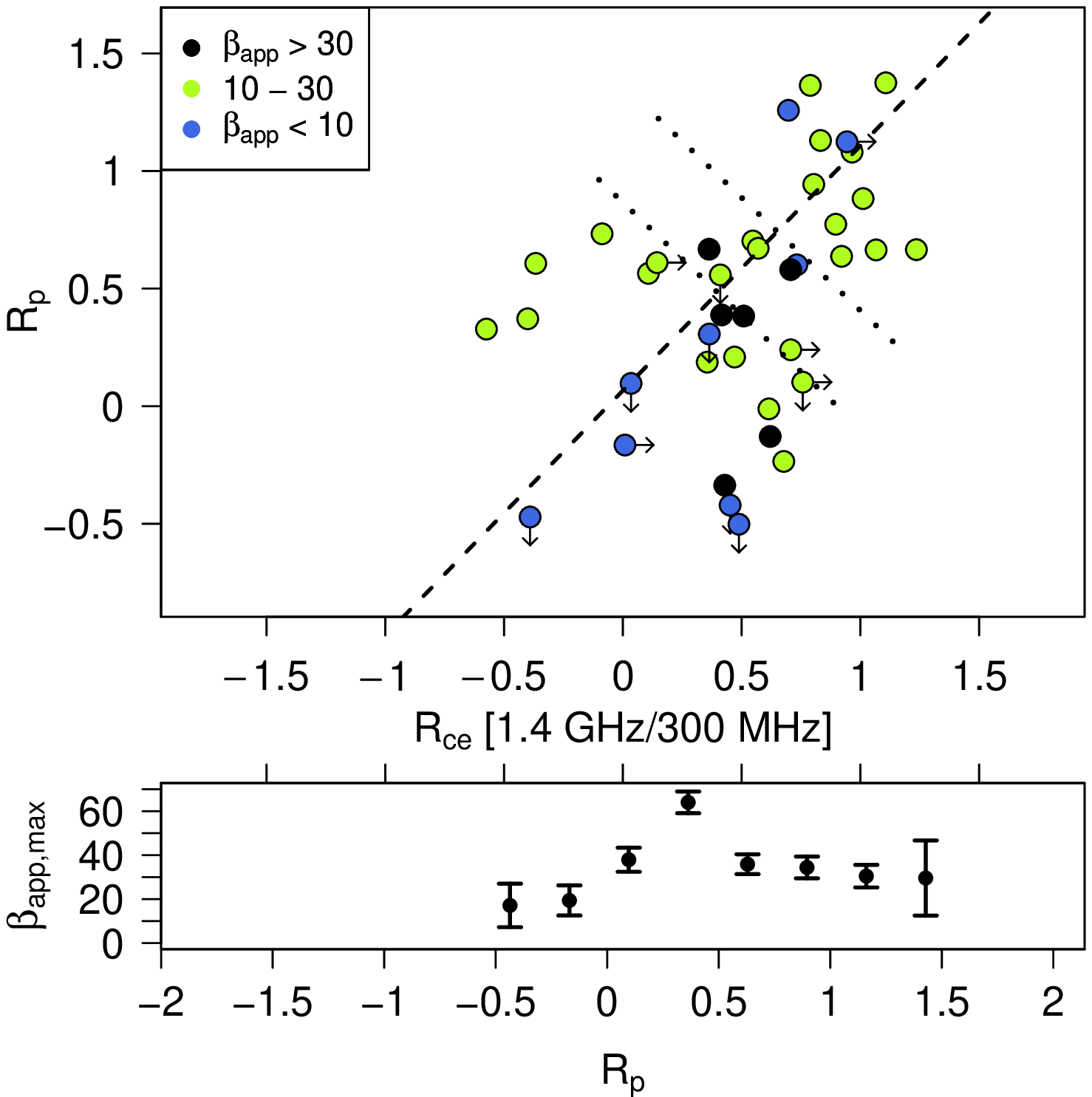}
\caption{\emph{Upper:} The same plot as Figure~\ref{ecsj}, for
  high-power sources (\lkin $>$ 10$^{45.5}$ erg s$^{-1}$) with
  measurements of maximum apparent jet speed (\bappns). Values are
  taken from \cite{jor01,jorstad05,kel04,lis09}, and \cite{kha10}. The
  appearance of low \bapp sources with high core and Compton dominance
  supports the interpretation that these sources are aligned very
  close to the line of sight. \emph{Lower:} Sources are binned along
  the shown correlation (running perpendicular to the dashed line in
  units of 0.25 of \rcens as illustrated by the two dotted lines), and
  the maximum \bapp is estimated. The increase in \bappmax from lower
  \rcens/\lgls towards a maximum, followed by a decrease towards the
  highest values is qualitatively as expected if \rce and \lgls
  increase with alignment.}
\label{bapp}
\end{figure}

The non-zero slope suggests EC emission in VHP sources; however, when
taken together the slopes in Figures~\ref{lgvsr} and \ref{echp} are
inconsistent with a simple EC scenario, as seen from
Table~\ref{table}. We have assumed that $\delta$ is the same for all
energies; if the radio-emitting plasma is slower, then the expected
slopes above will increase from the quoted ranges for both
figures. Assuming truly higher slopes for the VHP sources in
Figure~\ref{lgvsr}, concordance can be achieved with a general $p$=3
model with a slower $\Gamma$ in the radio.

\subsection{ The Test of  Superluminal Motions}
As shown by \cite{lis97}, a substantial number of sources
are expected to be seen at orientation angles smaller than that of
$\theta=1/\Gamma$ which maximizes the superluminal speed
$\beta_{app}$.  We therefore expect the highest \bapp values at \lgls
(\rce) values less than the maximum. Indeed if the highest \lgls
(\rce) sources are most aligned (within 1/$\Gamma$), their \bapp
values should be relatively small (noting that these may be
under-represented as their \bapp are difficult to measure).  In
Figure~\ref{bapp} (upper panel) the VHP sources are plotted as in
Figure~\ref{echp}, colored according to \bapp. The highest \bapp are
observed in the middle of the correlation, as expected. This is also
seen in the lower panel, where we have estimated the maximum \bapp in
bins of width 0.25 in \rcens, running parallel to the correlation
line.\footnote{The estimator for $\beta_\mathrm{app,max}$ is derived
  from the order statistic $Y_n$=max(\bappns) for a bin of size n
  using the estimator for the unknown upper bound of a uniform
  distribution $\beta_\mathrm{app,max}=Y_n(n+1)/n$ with variance
  $\beta_\mathrm{app,max}^2/(n^2+2n)$ \citep[e.g.][]{casella}.}

\subsection{Why is EC Only Apparent in Powerful Jets?}
What is apparently depicted in Figure \ref{echp} is that strong jets
at high \lkin are EC dominated, while lower \lkin strong jets are
not. However, the typical error in log \lkin of 0.7 is a significant
factor, as it may be the case that the restriction \lkin
$>$10$^{45.5}$ erg s$^{-1}$ is simply that which is high enough to
avoid any contamination with SSC sources.  Thus the true boundary
\lkin for the SSC to EC transition is probably lower than 10$^{45.5}$
(indeed, a few FSRQ of moderate \lkin are co-spatial in
Figure~\ref{echp} with the VHP sources). However, our findings imply
that (a) VHP sources are dominated by EC, and (b) many lower \lkin
strong jets appear at a similar (low) \lgls regardless of \rcens,
suggesting that EC is not important for some part of the population.

Which emission mechanism dominates is related to how fast the flow is:
assuming the gamma-ray emission of strong jets is inside the BLR or
MT, the comoving energy density of the external photon field is $
\Gamma^2U_\mathrm{ext}$.  The synchrotron energy density is
$U_\mathrm{s}=L_\mathrm{s}/(4\pi c^3 t^2_\mathrm{var} \Gamma^6)$, where
$t_\mathrm{var}$ is the observed variability timescale in hours, $L_\mathrm{s}$
the synchrotron luminosity, and we have assumed $\delta$=$\Gamma$. EC
will dominate over SSC provided $\Gamma^2 U_\mathrm{ext}>U_\mathrm{s}$, or for
$\G$ greater than a transition value
\begin{equation}
 \Gamma_\mathrm{tr}=16.2 \left(\frac{L_\mathrm{s}/10^{47}}{\left(t_\mathrm{var}/6\right)^2
   \left(U_\mathrm{ext}/10^{-4}\right)}\right)^\frac{1}{8}.
\end{equation}
Typical estimates of $U_\mathrm{ext}$ in the BLR and MT differ by
$\sim 100$ \citep[$U_\mathrm{ext, BLR} \approx 2.6\times 10^{-2}$ erg
  cm$^{-3}$ and $U_\mathrm{ext, MT} \approx 2.6\times 10^{-4}$ erg
  cm$^{-3}$;][]{ghi09_most_powerful}. In the MT we have
$\Gamma_\mathrm{tr, MT}$ $\sim$ 14.4, higher by a factor of
$100^{1/8}=1.8$ than in the BLR, where $\Gamma_\mathrm{tr, BLR}\sim$
8.1.

However, VLBI studies indicate that $\Gamma \gtrsim 10$ for most FSRQ
\citep[e.g.][]{jorstad05}; if the GeV emission site is in the BLR, it
is difficult to explain the lack of EC signature for some strong
jets. It has also been found that sources with higher \lkin produce on
the average faster superluminal motions \citep[][]{kha10}. The
connection between \lkin and $\Gamma$, \emph{and an emitting region in
  the MT}, can then explain a transition to EC at high \lkin: Strong
jets at lower \lkin are also slower and as long as
$\Gamma<\Gamma_\mathrm{tr}$ they are SSC emitters, exhibiting a
Compton dominance independent of radio core dominance. As \lkin
increases, $\Gamma$ also increases and once
$\Gamma>\Gamma_\mathrm{tr}$, the photon field of the MT begins to
dominate, producing a Compton dominance that increases with increasing
radio core dominance, as seen in Figure~\ref{echp}. If we adopt longer
$t_\mathrm{var}$, both $\Gamma_\mathrm{tr, BLR}$ and
$\Gamma_\mathrm{tr, MT}$ decrease, requiring that essentially all
strong jets are EC emitters, which our data does not support. On the
other hand, with a shorter $t_\mathrm{var}\sim$1 hr \citep[see
  e.g.][]{fos11} we obtain $\Gamma_\mathrm{tr}$ = 12.7 for the
BLR and $\Gamma_\mathrm{tr}$ = 22.5 for the MT, making either location
plausible.

\section{Conclusions}

The gamma-ray luminosities of \emph{Fermi}-detected radio-loud AGN have
been shown to depend strongly on both \lkin and the orientation angle.
We find the first collective evidence for external Compton emission in
high-power jets (\lkin $> 10^{45.5}$ erg s$^{-1}$): as can be seen in
Figure \ref{ecsj}, the Compton dominance of these sources increases
with radio core dominance, a measure of orientation. This requires
that the beaming pattern of the gamma-ray emission is more focused
than that of synchrotron, as is the case for EC scattering. A
confirmation that \lgls increases with decreasing orientation angle
comes from the fact that the apparent superluminal speeds are observed
to increase along the correlation to a maximum at moderate values of
\lglsns, before decreasing towards the highest values (Figure
\ref{bapp}), as expected since we anticipate a significant number of
sources at angles smaller than the maximum superluminal speed angle
$1/\Gamma$. The fact that strong jets of lower \lkin do not show an
increase of Compton dominance with alignment suggests that their
gamma-ray emission is due to SSC. Because more powerful jets appear to
be faster \citep{kha10}, a transition to EC at high \lkin can be
explained by a transition Lorentz factor above which the external
photons dominate.

\begin{acknowledgments}
EM and GF acknowledge support from NASA \emph{Fermi} grants NNX11AO15G and
NNX10AO42G, Swift grant NNX09AR04G and XMM grant NNX06AE92G. MG
acknowledges support from NASA ATFP grant NNX08AG77GS04 and \emph{Fermi}
grant NNX12AF01G. ML and the MOJAVE project is supported under National
Science Foundation grant 0807860-AST.
\end{acknowledgments}

\vspace{30pt}

\bibliographystyle{apj}
\bibliography{apj-jour,bib_be_letter,extra,add}

\end{document}